\documentclass[aps,prl,twocolumn,preprintnumbers]{revtex4}
\usepackage{graphicx}
\usepackage{amssymb}
\usepackage{amsmath}
\usepackage{times}
\usepackage{latexsym}
\usepackage{url}
\def\beq{\begin{equation}}
\def\eeq{\end{equation}}
\def\bey{\begin{eqnarray}}
\def\eey{\end{eqnarray}}
\begin{document}

\title{Bounds on Scalar Masses in Theories of Moduli Stabilization.}

\author{Bobby Samir Acharya}
\address{International Center for Theoretical Physics\\ Trieste, Italy \\
and\\ Department of Physics, KingÕs College London\\ London, UK}
\author{Gordon Kane}
\address{Department of Physics \& Michigan Center for Theoretical Physics, University of Michigan \\Ann Arbor, MI 48109, USA}
\author{Eric Kuflik}
\address{Department of Physics, Tel Aviv University\\ Tel Aviv, Israel }

\begin{abstract}

In recent years it has been realised that pre-BBN decays of moduli
can be a significant source of dark matter production, giving a
`non-thermal WIMP miracle' and
substantially reduced fine-tuning in cosmological axion physics.
We study moduli masses and sharpen the claim that moduli dominated the
pre-BBN Universe.
We conjecture that in any string theory
with stabilized moduli there will be at least one modulus field whose
mass is of order (or
less than) the gravitino mass. Cosmology then generically
requires the
gravitino mass not be less than about 30 TeV and the cosmological
history of the Universe is
non-thermal prior to BBN.
Stable LSP's produced in these decays can account for
the observed dark matter if they are `wino-like.'
We briefly consider implications for the LHC, rare decays, and dark
matter direct detection and point out
that these results could prove challenging for models attempting to
realize gauge mediation in string theory.

\end{abstract}
\maketitle

\section{Introduction}

String theories have moduli -- scalar fields with no classical potential and with Planck scale suppressed couplings to matter.
The expectation values of the moduli classically describe the size and configuration of the curled up
extra dimensions. In order to have a meaningful model describing phenomena
below the string scale, the moduli must be ``stabilized'', i.e., must have a
potential with a minimum that determines their value in the vacuum. Otherwise
observable coupling strengths and masses would not have meaningful values when
calculated from the theory, making it difficult to compare with data.

In a vacuum in which local supersymmetry is spontaneously broken there is
a goldstone fermion (goldstino) that becomes the longitudinal spin state
of the spin 3/2 gravitino, so that the gravitino is massive. The
gravitino mass -  $m_{3/2}$ - \textit{typically} sets the scale for all of the
scalar masses in the theory. Therefore, one generically expects moduli masses of
order $m_{3/2}$, far below the string scale. One of the aims of this paper is to sharpen
this last statement.

One expects that after inflation the Hubble scale $H_{I} \ge m_{3/2}$, and we will assume that this
is true throughout this paper.
Eventually, $H$ will decrease until $H \sim m_{\phi}$, the modulus mass, at which point the moduli
will begin to oscillate in their potential. One can estimate the energy
stored in the oscillating moduli, $\rho_{\phi}\sim m_{\phi}^{2}m_{pl}^{2}$
which has long been known \cite{Coughlan:1983ci,deCarlos:1993jw,Dine:1995uk} to be large and to dominate the energy
density of the Universe.  Since the moduli quanta are particles the Universe
is matter dominated during these oscillations, not radiation dominated. The
moduli ($\phi$) couple gravitationally to all other matter, with decay widths
$\Gamma_{\phi} \sim d_{\phi} m_{\phi}^{3}/m_{pl}^{2}$, where $d_{\phi}$ is a number of order unity
that is in principle calculable in any particular theory (see \cite{Moroi:1999zb,Acharya:2008bk} for examples).
Therefore, their lifetimes and decays can be estimated.

If the moduli decay after about $10^{-2}$ seconds after inflation ends, their
decay products will inject additional photons, hadrons and leptons during
Big Bang Nucleosynthesis (BBN), ruining its successful predictions. Lifetimes
shorter than about $10^{-2}$ seconds require moduli masses larger than about 30 TeV \cite{Weinberg:1982id,Ellis:1986zt,Kawasaki:1994af}.
If they do not
decay until after BBN we can estimate the energy stored in them from the energy density-to-entropy ratio
$\rho/s,$ which is approximately constant as the Universe cools.  Today,
$\rho/s$ is of order a few eV, but for the moduli it is of order $T_{osc}$,
the temperature of the Universe at the onset of the oscillations, which is
as large as $10^{11}$ GeV for moduli masses of order TeV. Since this is many orders of
magnitude larger than $\rho/s$ today, the moduli must decay
before BBN. This is the cosmological moduli problem \cite{Coughlan:1983ci,deCarlos:1993jw,Dine:1995uk}.
The only known way to avoid this result is to have a (tuned) period of
late inflation, called thermal inflation \cite{Coughlan:1984yk,Randall:1994fr,Lyth:1995ka,Moroi:1993mb,deGouvea:1997tn,Conlon:2007gk}  which could dilute the energy
density of the moduli fields. An interesting question is whether or not
thermal inflation might occur ``naturally'' in string theory, but we will leave
this for future work. In the following we will assume thermal inflation does
not occur. Models that generically produce thermal inflation have not been found.

In recent years, however, it has been realised that models with moduli which decay before BBN can
have virtues which are comparable to, or improvements upon, models which have a `thermal cosmological
history'. There can be a `non-thermal WIMP miracle' which is equally compelling as the thermal case \cite{Moroi:1999zb,NonThermMir}
and requires larger WIMP annihilation cross-sections.
Further, the entropy released from the moduli decays dilutes potential axion relic abundances and allows for much
less fine-tuned cosmological axion physics than is the case in a `thermal cosmological history', thereby relieving the tension
between cosmological bounds and GUT scale axion decay constants \cite{Fox:2004kb,Acharya:2010zx}. These virtues of a `non-thermal cosmological history' indicate
that the Cosmological Moduli Problem
is, perhaps, less of a problem and more likely part of a solution. Given the potential impact of such an indication, it is of importance to investigate
in more detail the claim that {\it generically the moduli masses will be of order $m_{3/2}$}.

We sharpen some of the existing arguments that realistic vacua arising from a
compactified string theory will, generically, have moduli or moduli-like fields
(such as hidden sector matter scalars, or axions), which dominate the
energy density of the Universe  prior to BBN. With moduli
stabilized, the moduli $F$-terms and those of the moduli-like fields contribute to
supersymmetry breaking, so the scalar goldstino will have significant moduli
components. We will argue that one or more moduli or moduli-like fields will have masses of
order the gravitino mass. These results can apply regardless of the value of the gravitino mass
and, hence, can give a strong constraint on the model of mediation of supersymmetry breaking.
We also discuss the moduli spectrum in a variety of different classes of vacua in which moduli stabilization is
fairly well understood and demonstrate that there is always a modulus-like field which dominates the cosmic energy density
prior to BBN. These latter results are described in Appendix B.

We have thus given further evidence that,
from the string theory point of view, the Universe is expected to have a partly non-thermal history
-- an important claim for considerations of cosmology, especially dark matter. The gravitino mass
is required to be greater than about 30 TeV implying that gauge mediated supersymmetry breaking will be difficult to realise in
a phenomenologically consistent string vacuum.

The next section describes the main results which are based on general considerations of supergravity and follow the
earlier approach of \cite{Denef:2004cf,GomezReino:2006dk}. 
The following section describes the main physics implications of the result.

\section{Supergravity and Moduli Masses}

The conjecture: In any string/$M$ theory vacuum with observationally
consistent energy density of the Universe, there exists at least one modulus-like field whose
mass is such that it dominates the vacuum energy up to the BBN era.

Here, the term modulus-like refers not only to the geometric moduli fields of
string theory, but includes other scalar fields whose couplings to Standard
Model particles are suppressed by a high scale such as the Grand Unification
scale or Planck scale. Examples include axions and other hidden sector fields.

Typically, as we will see, the conjecture follows from the fact that the
masses of these moduli-like fields are of order, or less than, the gravitino mass $m_{3/2}$; in fact
this would be a more general version of the conjecture.
As we will see below, there can be examples in which the relevant
mass scale is much less than $m_{3/2}$ because of ``large volume'' effects in the extra
dimensional theory \cite{largevolume}. These examples do not violate either conjecture.

Since the moduli fields are stabilized by assumption, they have
non-trivial potentials and will most likely participate in supersymmetry
breaking. If this is true, i.e., at least one moduli has a non-vanishing $F$-term, then one can make progress towards proving the
conjecture by considering the scalar potential of the effective supergravity
theory. (String vacua in which the 4-D supergravity approximation is not valid may be beyond the scope this paper). 

The scalar potential in a supergravity theory in four dimensions can be
written in terms of a single real function $G$. In terms of the Kahler
potential and superpotential $G = K + m_{pl}^{2} ln (W\bar{W}/m_{pl}^{6} )$,
though we will only consider $G$ here. $G$ is taken to have mass dimension two
and all scalar fields are taken to be dimensionless in order to exihibit
correctly the scaling of various operators with the reduced Planck mass $m_{pl}$. The
scalar field potential, assuming that the $D$-term contributions are negligible is

\begin{equation}
V = m_{pl}^{2} e^{G/m_{pl}^{2}} \left(  G^{i} G_{i} -3 m_{pl}^{2} \right)
\end{equation}

Critical points of $V$ satisfy
\begin{equation}
\left<  \nabla_{i} V \right>  = \left< m_{pl}^{-2} G_i V + e^{G/m_{pl}^{2}} \left( G_{i} + G^{k} \nabla_{i} G_{k} \right) \right>  = 0 .
\end{equation}
where subscripts on $G$ denote derivatives with respect to the moduli fields $\phi_i$ or their conjugates $\phi^*_{\bar{i}}$.

One can calculate the complex scalar mass matrix in the vacua which satisfy the above conditions
and, additionally, have zero cosmological constant. Assuming the kinetic terms have been properly normalized, the mass matrix for all the scalar fields in the theory is
\begin{equation}
\begin{array}
[c]{ccl}%
M^{2}_{i \bar{j} } & = & e^{G/m_{pl}^{2}} \left(  \nabla_{i} G_{k}
\nabla_{\bar{j}} G^{k} - R_{i\bar{j} k \bar{l}} G^{k} G^{\bar{l}} + G_{i
\bar{j}} \right) \\
M^{2}_{i j } & = & e^{G/m_{pl}^{2}} \left(  2 \nabla_{i} G_{j} + G^{k}
\nabla_{i} \nabla_{j} G_{k} \right)
\label{massmatrix}
\end{array}
\end{equation}
Notice that there is no factor of $m_{pl}^{2}$ because the quantities in the
brackets all have mass dimension two.

Since we are only interested in minima of
the potential, the mass matrix is positive definite by assumption. Hence, we use the theorem 
that its
smallest eigenvalue, $m_{\min}^{2}$, is less than $\xi^{\dagger} M \xi $ for
any unit vector $\xi$.
Extending the work of \cite{Denef:2004cf,GomezReino:2006dk}, we take
$ \xi = (G^{\bar{j}} \;\; c\, G^{j} ) / \sqrt{3(1 + |c|^{2})} $ for $c
\in\mathbb{C}$, which is aligned in the (moduli components of the) two sGoldstino directions $\eta= G^{i} \phi_{i}$ and $\bar{\eta} = G_{i} \phi^{i}$.
This gives a one (complex) parameter class of constraints on the
upper bound of the lowest mass eigenvalue
\begin{equation}
\begin{array}{ccl}
m_{\min}^{2} &\le&
\frac{1}{3(1 + |c|^{2})} \left(
\begin{array}{cc}
 G^i & c^{\dagger }G^{\bar{i}}
\end{array}
\right)\left(
\begin{array}{cc}
 M_{i \bar{j}}^2 & M_{ij}^2 \\
 M_{\bar{i} \bar{j}}^2 & M_{\bar{i} j}^2
\end{array}
\right)\left(
\begin{array}{c}
 G^{\bar{j}} \\
 c G^j
\end{array}
\right) \\
 &\le& m_{3/2}^{2} \left( 2 \dfrac{|1 - c|^{2}}{1+|c|^{2}} +
\mbox{Re} \{ \dfrac{2 c}{1+|c|^{2}} {\dfrac{u}{m_{pl}^{2}}} \} - {\frac{r
}{m_{pl}^{2}}} \right)
\label{MminC}
\end{array}
\end{equation}
where $u \equiv\frac{1}{3} G^{i} G^{j}  G^{k} \nabla_{i} \nabla_{j} G_{k} $, $r
\equiv\frac{1}{3} R_{i\bar{j} k \bar{l}} G^{i} G^{\bar{j}} G^{k} G^{\bar{l}} $
and $m_{3/2}^{2} = m_{pl}^{2} e^{G/m_{pl}^{2}}$. $r$ is the holomorphic
sectional curvature of the scalar field space, evaluated in the sGoldstino directions
in field space. We have extended the previous work to include the effects of the curvature $r$ as well as $u$.

To understand the constraint given by Eq. (\ref{MminC})  we rewrite this equation by taking $u = |u|
e^{i \theta_{u}} $
\begin{equation}
m_{\min}^{2} < m_{3/2}^{2} \left(  2 - 2 \alpha\cos\theta+ \alpha{\frac{|u|
}{m_{pl}^{2}}} \cos(\theta+\theta_{u}) - {\frac{r }{m_{pl}^{2}}} \right)
\end{equation}
for any $\alpha \equiv \frac{2 |c|}{1+|c|^2} \in[0,1]$ and $\theta\in[0,2\pi] $.

It therefore follows that
\begin{equation}
m_{\min}^{2} < m_{3/2}^{2} \left(  2 - {\frac{r }{m_{pl}^{2}}} \right)
\label{rlimit}
\end{equation}
and for $u\in\mathbb{R}$
\begin{equation}
m_{\min}^{2} < m_{3/2}^{2} \left(  \min\{{\frac{u }{m_{pl}^{2}}},4-{\frac{u
}{m_{pl}^{2}}} \} - {\frac{r }{m_{pl}^{2}}} \right)  .
\label{ulimit}
\end{equation}
So, as long as $|r|, |u| \leq\mbox{O} (m_{pl}^{2} )$, the upper limit on the
lightest modulus mass is of order the gravitino mass $m_{3/2}$.  (See Appendix A for a simple
model illustrating the bound given by Eq. (\ref{rlimit}) and  Eq. (\ref{ulimit})).

In fact, for
geometric moduli, $r$ is typically of order $m_{pl}^{2}$, and thus there will generically be at least one moduli with mass $\lesssim m_{3/2}$. 
In Appendix B we systematically discuss the moduli masses in all known (at least to us)
examples where moduli stabilization is well understood. It is demonstrated that all of these
examples have $r \sim m_{pl}^2$ and a modulus or modulus-like field whose mass is less than, or of order $m_{3/2}$

\section{Non-generic Possibilities}
One can discuss under what non-generic conditions moduli, or moduli-like fields, will
\textit{not} dominate the cosmic energy density prior to BBN. One possibility is that {\it all} moduli that have mass order $m_{3/2}$
have significant mixing with charged (under the SM gauge groups or
other gauge groups) scalar fields. Then the lightest eigenvalue
given by Eq. (\ref{ulimit}) and  Eq. (\ref{rlimit}) can have
significant charged matter components and quickly thermalises due to
it's couplings to gauge fields and
matter. However, mixing between moduli and matter
is proportional to vevs of the matter fields or their $F$-terms and
are usually suppressed. It would be difficult to arrange for all moduli
to have such large couplings to matter fields. But if any stabilized modulus field (or linear combination) does not mix, our results will hold.  Another possibility is to have moduli stabilization
unrelated to supersymmetry breaking and a stabilization mechanism that gives
all moduli very large masses, but as stated earlier, non-trivial potentials that stabilize moduli will generally break supersymmetry.
 A third is to have $r$ or $u$ extremely large
(and negative), e.g. $|r| \gg (30\mbox{ TeV}/m_{3/2})^{2}m_{pl}^{2}$, or have very large kinetic terms so that the mass matrix (Eq. \ref{massmatrix}) receives large scaling factors when the kinetic terms are properly normalized.
At present such non-generic cases are not excluded, but any proposed model has to explain why they
might occur. A final possibility is to have moduli of order $m_{3/2}$, but not oscillate in the early universe, for example, see "moduli trapping" \cite{moduliesp,swatson} or
"string gas cosmology" \cite{Brandenberger:1988aj,Battefeld:2005av}.

Part of the difficulty in trying to make such a model is that there is always
at least one axion field present in four dimensional string theory vacua;
usually, there are many axions \cite{Witten:1984dg,Kim:1986ax,Svrcek:2006yi,Arvanitaki:2009fg}. Some of these axions will
be much lighter than the moduli, which are assumed to be heavy enough to be
cosmologically irrelevant in this part of the discussion. This is because
axions only gain mass via non-perturbative effects. One can calculate the
relic abundance of such an axion today, as a function of its mass, $m_{a}$,
decay constant, $f_{a}$ and initial displacement $\theta_{a}$ It is given by \cite{Fox:2004kb,Acharya:2010zx}
\begin{equation}
\label{thermal}\Omega_{a}\,h^{2} = 0.06\;\left(  \frac{f_{a}^{2}}{M_{GUT}^{2}%
}\,\right)  \left(  \frac{m_{a}}{10^{-20}\text{ eV}} \right)  ^{1/2}\langle\theta
_{a}^{2}\rangle\,
\end{equation}

With a GUT scale decay constant, the axion lifetime extends into the BBN era
if $m_{a}\leq100$ GeV, so the above formula implies that one would have to
tune the model such that there are no axions with masses between $10^{-20}$ eV
and 100 GeV!

\section{Non-thermal cosmological history from string theory}

The arguments based on the moduli mass matrix imply that $m_{\phi}\le
O(m_{3/2}),$ and that $m_{3/2}$ must be of order $30$ TeV or larger to not conflict
with BBN predictions or the observed late energy density \cite{deAlwis:2010sw}.  The moduli couple essentially
universally to every Standard Model particle and their superpartners.  Some moduli decays
might be helicity suppressed, but decays to scalars are all present at full
strength.  These decays generate huge entropy, which significantly dilutes any dark
matter particles that might have been present before the moduli decay. \ It is
sufficient if only one such modulus has mass of order $m_{3/2}$ though typically
many do.  {\it Thus thermal freezeout relic densities of dark matter are not
relevant to present cosmology in string models where moduli are present.}  At the same time, about a quarter \cite{Moroi:1999zb,Acharya:2008bk} of all moduli decays will be to
superpartners, and every superpartner will have a decay chain with a lightest
superpartner (LSP) at the end, so a large number of LSP dark matter particles
will be generated and provide a dark matter candidate if they are stable.  In practice, the
number density of LSP's from moduli decay is large compared to the relic density, large enough
for LSP's to annihilate,
The typical temperature after the moduli decay is of order 10 MeV,
but the number density decreases as the Universe expands,
approaching an attractor solution of the Boltzmann equations (when the number density is
too small for annihilation to take place)
and not a freeze-out (which occurs when the LSP's fall out of equilibrium).
Surprisingly, one still finds a ``WIMP miracle'', where the relic number density
is still given in terms of the Hubble parameter - a cosmological parameter, but
now evaluated at the moduli decay temperature rather than a freezeout
temperature - and a particle physics annihilation cross-section for the LSP's (appropriately averaged) \cite{NonThermMir}
\begin{equation}
N_{LSP}\approx H(T \sim10\text{ MeV})/\left\langle \sigma_{annih}^{LSP}\text{v}\right\rangle .
\end{equation}

All steps of this calculation have been carried out in the example of
$M$ theory compactified on a manifold of $G_{2}$ holonomy, including the moduli
stabilization, calculation of the moduli masses and decays and the entropy
generated, etc (see \cite{g2mssm}). Importantly, in
order to obtain about the right relic density, the LSP must be a wino or
wino-like particle, with a large annihilation cross section of about $3\times10^{-24}%
{\mbox{cm}}^{3}/\sec.$ Such an LSP naturally arises in the $G_{2}$-MSSM \cite{Acharya:2006ia,Acharya:2007rc}, where the tree
level gaugino masses are suppressed by the supersymmetry breaking mechanism
to be of order the gaugino masses from the anomaly mediation contribution.
Thus the non-thermal history and a wino LSP go together and give a
consistent picture for dark matter from the compactified string theory. 

One might wonder if the moduli ``reheat temperature'', effectively the
temperature generated by the moduli decay, could be above the thermal
freezeout temperature so that the thermal history could finally take over.
Unfortunately, the associated temperature is too small for a thermal history to develop. To see this let $T_{RH}\sim10$ MeV for
$m_{3/2}\sim 30$ TeV, which follows from $T_{RH} \sim \sqrt{\Gamma_{\phi}m_{pl}} \sim {m_{3/2}^{3/2} \over m_{pl}^{1/2}}$.
\ Then if $m_{3/2}$ were larger by even an order of
magnitude, $T_{RH}$ would grow by a factor of order $(10)^{3/2}\sim30$, so it
would still be small compared to the usual thermal freezeout temperature $T_{fr}\sim $ few
GeV.

Thus very generally string theories with stabilized moduli having multi-TeV
scale masses (or lighter) will have a non-thermal cosmological history, and a
relic density of wino-like dark matter generated by moduli decay rather than
thermal freezeout is the prefered solution with LSP dark matter.
\ A fine-tuned period of late inflation
may allow a way to evade this generic conclusion for some theories, but an
inflaton with just the right properties must be found in such a theory.

\section{Gauge Mediation Supersymmetry Breaking}

Our results suggest that any approach to supersymmetry breaking that originates in a string theory
with moduli that has a gravitino  mass less than about 30 TeV will have
the problems described above, the moduli and gravitino problems.   Thus, one would conclude that gauge mediated
supersymmetry breaking, which typically has a much lighter gravitino and therefore light moduli, does not generically arise if our universe is described by a compactified string
theory with stabilized moduli.

\section{Heavy Scalars, Light Gauginos, LHC, and Rare Decays}

All superpartner masses in gravity mediated supersymmetry breaking are proportional to
$m_{3/2}$. \ Scalar masses generically will have values about equal to
$m_{3/2}$, but gaugino \ masses are often suppressed, usually because the main
source of supersymmetry breaking does not couple at tree level with the
gaugino \ masses, and they are zero in the supersymmetry limit. \ Several
phenomenological consequences follow from these properties of generic
string theory vacua. \

At LHC the scalar superpartners should not be observed directly. \ The gauginos, \ in
particular the gluino, the lightest two neutralinos (including the LSP), and
the lighter chargino can be observed. 
Because the squark masses at the weak scale are given by
running from the gravitino mass scale there are effects on gluino branching
ratios even though the scalars cannot be directly observed, with a large BR of
a gluino into top quarks (so gluino pairs often have 4 tops per event), and in general
considerably larger BRs to channels with final b quarks, leading to rich LHC physics \cite{Acharya:2009gb,FKRN,Giudice:2010wb}. Any effect on decays
or moments of quarks and leptons that can only occur from loops should not
differ from its Standard Model value significantly; in particular $g_{\mu}-2$,
$B_{s}\rightarrow\mu^{+}\mu^{-}$, a charge asymmetry from $B_s$-mixing like-sign dimuons, and
other effects should all take on their SM values.
Finally, heavy stops, of order 30 TeV, raise the Higgs above 
the tree-level MSSM bound, consistent with a 125 GeV Higgs boson.

Some predictions depend on how the $\mu$ problem is solved in string theory,
which is rather poorly understood. \ On the one hand the $\mu$ term in
the superpotential must vanish so $\mu$ does not have a string scale value,
presumably because of a symmetry.  But $\mu$ and the supersymmetry breaking
$B\mu$ terms must be non-zero so the symmetry that protects $\mu$ must be
broken. When that symmetry is broken doublet-triplet splitting must be
preserved, the proton must not decay too rapidly, and the LSP must have a
lifetime longer than about $10^{26}$ sec \cite{LSPLifetime}. If the resulting value of $\mu$
is similar to the value of $M_{2}$ that gives the wino mass, then there is a
light higgsino that mixes into the LSP and the light chargino. A higgsino
admixture in the neutralino mass matrix gives an off diagonal term that
dilutes the wino, and necessarily mixes in some bino as well. The main
observable affected by the small $\mu$ is actually the scattering cross
section in direct detection experiments. For a pure wino LSP the cross-section for
LSP--proton scattering is below $10^{-46}{\mbox{ cm}}^{2}$, while with a higgsino
admixture it can get as large as a $few\times10^{-44}{\mbox{ cm}}^{2}$. Thus future
direct-detection measurements will determine the allowed higgsino mixture in the wino, and
approximately measure the value of $\mu.$

\section{Conclusions}

We have argued that if our universe is described by a
compactified string theory then the presence of stabilized moduli would likely
imply that the cosmological history is non-thermal before BBN. In
particular, dark matter can be produced from moduli decays
and generically has to be wino-like in order to have a consistent abundance.
The analysis
that leads to these results sharpens the arguments that there is
always a modulus with mass of order the gravitino mass or less in such
theories. These plus cosmological considerations emphasise
some difficulties in realizing gauge mediated supersymmetry breaking in
string theory. We also described an upper limit on the gravitino mass of order several hundred TeV.
The appendix contains a study of the known examples of string theory vacua with stabilized
moduli and shows they agree with the results above.

\section{Acknowledgements}

We appreciate helpful conversations with  Shanta de Alwis, Konstantin Bobkov, Frederik Denef,
Michael Douglas, Daniel Feldman, Piyush Kumar, Aaron Pierce, and Fernando Quevedo.  We would also like to thank C. Scrucca and J. Louis for pointing out an error in the previous version of this article. B.A.
is grateful to the University of Michigan
Physics Department and MCTP for support, and E.K. is grateful to the String Vacuum Project for travel support and for a  String Vacuum Project Graduate Fellowship
funded through NSF grant PHY/0917807. This work was supported by the DOE Grant \#DE-FG02-95ER40899.

\appendix
\section{}

For pedagogy we begin with a few toy (non-string theory) examples to demonstrate the validity of the above results.

Consider the Polonyi model, which has one field $\phi$ with $G$ given by
\begin{equation}
G= m_{pl}^{2} \phi\bar{\phi} + m_{pl}^{2} \log \left|\frac{\mu^{2}}{m_{pl}^{2}} (\phi- \beta) \right|^2
\end{equation}
where $\beta = \sqrt{3}-2$. The vacuum expectation value of $\phi$ is given by%
\begin{equation}
\left<  \phi\right>  = \left<  \bar{\phi} \right>  = \sqrt{3} - 1
\end{equation}
and $u$ and $r$ are
\begin{equation}
u=2\sqrt{3}m_{pl}^{2},\text{ \ }r=0
\end{equation}
The two eigenvalues of the mass matrix are
\begin{equation}
\begin{array}[c]{ccl}
m_{1}^{2} & = & m_{3/2}^{2} ( 2\sqrt{3} ) = m_{3/2}^{2}{\frac{u }{m_{pl}^{2}}} \\
m_{2}^{2} & = & m_{3/2}^{2} ( 4- 2\sqrt{3}) = m_{3/2}^{2}(4 - {\frac{u }{m_{pl}^{2}}})
\end{array}
\end{equation}
in agreement with our general result Eq. (\ref{ulimit}). The bounds are reached since the eigenvectors are given by the sGoldstino directions.

This provides as an illustration of how one might avoid our result by going to a non-string theory withour moduli stabalization, by creating a model with large $r$ and have scalar masses much heavier than the gravitino mass. Following \cite{Kitano1,Ibe:2006rc}, we add a higher dimensional operator to the above Kahler potential
\begin{equation}
G = m_{pl}^{2} \phi\bar{\phi} +  m_{pl}^{4} \frac{(\phi\bar{\phi})^2}{\Lambda^2} + m_{pl}^{2} \log \left|\frac{\mu^{2}}{m_{pl}^{2}} (\phi- \frac{1}{\sqrt{3}}) \right|^2.
\end{equation}
with the new scale $\Lambda \ll m_{pl} $ .
There is a minimum of this potential where $\phi$ has a small vacuum expectation value. To leading order in $\frac{\Lambda}{m_{pl}}$
\begin{equation}
\left<  \phi \right>  = \left<  \bar{\phi} \right>  \approx \frac{\Lambda^2}{2 \sqrt{3} m_{pl}^2} .
\end{equation}
In this vacuum $r = -12 \frac{m_{pl}^4}{\Lambda^2} $, and one find after diagonalizing the scalar mass matrix that
\begin{equation}
m_{1}^{2} = m_{2}^{2} = 4 \mu^2 \frac{m_{pl}^2}{\Lambda^2} = 12 \frac{m_{pl}^2}{\Lambda^2} m_{3/2}^{2} = -\frac{r}{m_{pl}^2} m_{3/2}^{2} \gg m_{3/2}^{2}.
\end{equation}
Thus the scalar masses are at the bounds given by Eq. (\ref{rlimit}) and are much heavier than the gravitino mass. The field here has explicit couplings that are not moduli-like and if one tries to embed this model in string theory new problems arise -- see Ex. 6 \cite{Dudas:2008qf}.

\section{}

We now wish to examine examples we know of of string theory models in which all
moduli are stabilized to gain further insight.

\textbf{Ex. 1} Simple KKLT Model

This example \cite{kklt} has all complex structure
moduli stabilized by fluxes and a single Kahler modulus and axion stabilized
by non-perturbative corrections. The vacuum energy is tuned by adding what
amounts to a D-term potential. We didn't consider D-terms above, but we can
incorporate them into the discussion based on these examples. Both the Kahler
modulus and the axion obtain masses of order $20.\times m_{3/2}$ in this simple model. Here $r \sim 0$, but 
the supergravity theory is not supersymmetric, due to the explicit D-term breaking. 
This scenario allows for a lighter gravitino  $m_{3/2}\sim 1$ TeV, but the moduli will still 
dominate the cosmic energy density for times up to BBN and beyond.

 The non-thermal cosmology of mirage-mediated supersymmetry breaking in the KKLT context was discussed  \cite{Nagai:2007ud}. Late decay of moduli produce an abundance of Bino-like LSPs, which the annihilate rapidly through the pseudo-scalar Higgs resonance.

\textbf{Ex. 2} LARGE Volume IIB models

These examples \cite{largevolume} have complex structure moduli stabilized by fluxes and Kahler
moduli stabilized by perturbative corrections. The vacuum energy is, naively,
negative, though there might be mechanisms which generate the necessary
positive contributions. The basic LARGE volume model has two Kahler moduli
$\tau_{b}$ and $\tau_{s}$. In the vacuum, $m_{3/2} \sim{\frac{m_{pl} }{V}}$,
where $V$ is the volume of the extra dimensions (divided by $l_{s}^{6}$). The
masses of the moduli are given by $m_{\tau_{b}} \sim{\frac{m_{3/2} }{V^{1/2}}%
}$ and $m_{\tau_{s}} \sim m_{3/2}\log{m_{pl}/m_{3/2}}$. $\tau_{b}$ is much
lighter than the gravitino and $\tau_{s}$ is an order of magnitude larger. Note that
the suppression of the $\tau_b$ mass in this case can be shown to be from a direct
cancellation in Eq. (\ref{rlimit}) as $r/m_{pl}^2 = 2 - O(1/V)$.
In these models, unless $V \leq10^{9}$, $\tau_{b}$ generically suffers from the
cosmological moduli problem. In all cases, the early Universe is dominated by
moduli oscillations. More recently, it has been realised that, by adding a
third Kahler modulus, the observable sector supersymmetry breaking masses
are suppressed relative to the gravitino mass, requiring $10^{8}\mbox{ GeV}\leq
m_{3/2} \leq10^{11}\mbox{ GeV}$ \cite{Blumenhagen:2009gk}. Again, in all such cases, $\tau_{b}$ dominates the
pre-BBN cosmic energy density.

\textbf{Ex. 3} $M$ theory and Type IIA flux Vacua

These examples \cite{Acharya:2002kv,Villadoro:2005cu,Acharya:2005ez,DeWolfe:2005uu,Acharya:2006ne} use fluxes to stabilize all the moduli. All these vacua have a
negative cosmological constant and it seems difficult to add additional
sources which could change that. The moduli masses are all of order the
gravitino mass.

\textbf{Ex. 4} $M$ theory on Manifolds of $G_{2}$ holonomy without flux.

These examples \cite{g2mssm} are based upon the idea, which goes back to Witten and others \cite{WA},
that strong dynamics in the hidden sector generates a potential which breaks
supersymmetry and generates a hierarchically small scale (related to the weak
scale). In the M theory context it has been shown that, additionally, the
potential generated by such hidden sector dynamics can stabilize \textit{all}
the moduli fields. The minimum of the potential has positive energy. The
moduli spectrum for these examples has been studied in detail in \cite{g2mssm}. All
moduli but one have masses of order $m_{3/2}$, the remaining one having a
somehwat larger mass. Hence, again, the moduli dominate the early cosmological history but decay before BBN.

\textbf{Ex. 5} Type IIB flux vacua with non-perturbative effects

These examples \cite{osu} apply the ideas of \cite{Acharya:2006ia,Acharya:2007rc} to stabilize all Kahler moduli and
obtain a vacuum with positive vacuum energy self consistently. These examples
all have moduli whose masses are of order $m_{3/2}$ and hence will dominate
the early Universe.

\textbf{Ex. 6} Gauge Mediation in String theory?

In gauge mediation, the gravitino mass is relatively low and can be as
small as an eV. Generically one expects that there are moduli whose
masses are comparable to $m_{3/2}$. Since their lifetimes are so long,
these moduli will
dominate the Universe for many years and will not be able to reheat it
to a temperature high enough for BBN to start (see \cite{GG} for a
discussion on BBN constraints). Usually, when one
considers gauge mediation, one implicitly assumes that moduli can be
decoupled from the gravitino mass scale and then ignored, but our
results indicate that such assumptions are perhaps too strong.
Attempts at realizing a supergravity model
derived from string theory with both moduli stabilization and gauge
mediation
are described in \cite{Dudas:2008qf}, based on earlier works of
\cite{Kitano1,Ibe:2006rc,Lalak:2008bc}.
These models essentially couple a Type IIB Kahler moduli sector to a
gauge
mediation model which is assumed to arise from a configuration of branes
on
the Calabi-Yau of the sort described in
\cite{Diaconescu:2005pc}.
The authors of
\cite{Dudas:2008qf} explain that it is quite difficult to find a model
in which gauge
mediation effects are not overcome by those of gravity mediation, when
the
cosmological constant is tuned to zero. In any case, if one examines the
moduli masses in those examples one finds that the moduli whose masses
are
dominated by $D$-terms have masses much larger than $m_{3/2}$, but those
whose
masses are dominated by $F$-terms have masses of order $m_{3/2}$.
Therefore, generally one has
moduli which lead to a non-thermal comological history.

\end{document}